# DYNAMICS OF SPINING PARTICLES IN A HOMOGENEOUS SPACE WITH ROTATION AS POSSIBLE MECHANISM OF THE INERTIAL MASS FORMATION AND INTERPRETATION OF QUANTUM UNCERTAINTY PRINCIPLE IN GENERAL RELATIVITY

V.G.Krechet [1], V.B.Oshurko [1,2], A.E.Kisser [1]

[1] *Moscow State Technological University "Stankin"*
[2] *Prokhorov General Physics Institute of Russian Academy of Sciences*

Abstract. The dynamics of particles with intrinsic angular momentum (spin) described by the Dirac equation is considered in a homogeneous space with rotation in the presence of a homogeneous vortex gravitational field. The effects of the interaction between the spin of Dirac particles and the vortex gravitational field, as well as a possible mechanism for the appearance of an inertial mass, which is inextricably linked, as shown, with the spin of particles, are revealed. A geometric interpretation of the uncertainty principle is given..

Recently, the Gödel metric has attracted some interest [1-3]. It is known that the Gödel metric is a solution to the Einstein equation for a space filled with non-interacting rotating particles. It was originally obtained as a cosmological solution, but was not accepted for two reasons. Firstly, the Gödel metric does not predict the expansion of the universe and, secondly, it gives closed time-like geodesics, which means a violation of the causal relationship. However, the modern (Copenhagen) interpretation of quantum mechanics also assumes, in a certain sense, a violation of the causal relationship in such phenomena as the collapse of the wave function or as a consequence of the restrictions imposed by the uncertainty principle. Given that the spin in its manifestations is largely equivalent to the moment of rotation of the particle, it can be assumed that the space-time metric in a system of such particles will be similar to the Gödel metric. It is interesting to consider the dynamics of particles with spin, described by the Dirac equation, in space-time with a Gödel-type metric (i.e., in a homogeneous space-time with rotation, or, in other words, in the presence of a homogeneous vortex gravitational field) and possible effects of interaction of space rotation and spin. As will be shown, this will allow us to propose an interpretation of quantum mechanics based on general relativity.

In present work the dynamics of spinning particles (described by the Dirac equation) in a homogeneous space-time with rotation is analyzed. Possible effects of interaction between the rotation of space and spin of particle are considered.

One of the simplest metrics describing a homogeneous space-time with rotation is a two-parameter metric, which is the generalization of the Gödel metric of a homogeneous rotating cosmological model [4]

$$ds^2 = dx^2 + ke^{2\lambda x}dy^2 + dz^2 + 2e^{\lambda x}dydt - dt^2 . \quad (1)$$

Here, time $t$ has the dimension of length ($cm$) and is related to world time $t_k$ (c) by the relation $t = ct_k$, and the parameter $\lambda$ determines the angular velocity of rotation of the cosmological model around the OZ axis, the constant $k$ is the causality parameter. When $k < 0$ the causality is broken: at least one closed time-like geodesic passes through each point of space-time. When $k > 0$, then there are no closed time-like geodesics, and there is a usual causal structure in space-time.

This situation is illustrated in Fig. 1. In this figure the example of timelike geodesics for various values of $k$, obtained by numerical simulation is presented. It can be seen from Fig. 1 that at $k < 0$, for example, at $k = -0.75$, the timelike geodesic, leaving the point (0, 0) at $t = 0$, then returns two more times to the moment $t = 0$ at points with other values of the $y$ coordinate.

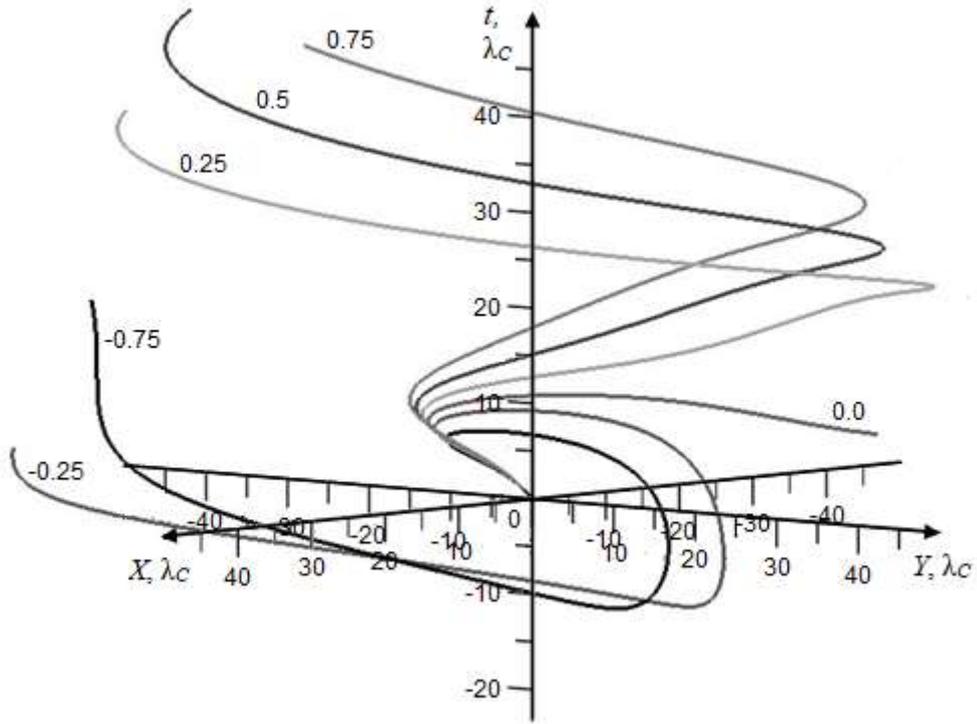

Fig. 1. World lines of a particle in Cartesian coordinates x and y (the time axis (*ct*) is vertical, $\lambda_C$ is the reduced Compton wavelength)

Metric (1) is the closest generalization of the Gödel metric [5] for a homogeneous (!) stationary rotating cosmological model

$$ds^2 = dx^2 - \frac{1}{2}e^{2\lambda x}dy^2 + dz^2 + 2e^{\lambda x}dtdy - dt^2 \qquad (2)$$

It is easy to see that the Gödel metric (2) is a special case of metric (1) at $k=-1/2$. Using the co-moving frame of reference for the spinor particles and the observer, their 4-velocity $V^k$, normalized to unity, can be represented as:

$$V^i = \tau^i = (0,0,0,1); \quad \tau_i = (0, e^{\lambda x}, 0, -1); \quad \tau^i \tau_i = -1. \qquad (3)$$

In this frame of reference, the angular velocity of rotation $\omega^i$ of space, determined by the formula

$$\omega^i = \frac{1}{2}\varepsilon^{iklm}\tau_k \tau_{l,m}, \qquad (4)$$

will be described in this case by the expression

$$\omega^i = \frac{\lambda}{2\sqrt{k+1}}\delta^i_3; \quad \omega = \sqrt{\omega_i \omega^i} = \frac{\lambda}{2\sqrt{k+1}}. \qquad (5)$$

Here the angular velocity vector is directed along the OZ axis, hence, the rotation axis is the OZ axis, and the value of the angular velocity $\omega$ is a constant value. Moreover, since the space-time with the metric (1) is homogeneous, then the axis of rotation passes through each point of this space.

On the other hand, in a homogeneous space with metric (1) there is a homogeneous vortex gravitational field, which is determined in the general case by a 4-dimensional curl of the field of tangent tetrads $e^k_{(a)}(x^i)$ [6]

$$^{\upsilon}\omega^i = \frac{1}{2}\varepsilon^{iklm}e_{k(a)}e^{(a)}_{l,m}. \quad (6)$$

Here $e^k_{(a)}$ are tetrad coefficients, $^{\upsilon}\omega^i$ is the angular velocity of the tangent tetrad, which is a kinematic characteristic of the vortex gravitational field. The axial 4-vector $^{\upsilon}\omega^i$ determines the density vector of the intrinsic angular momentum $S^i(g,x^i)$ of the vortex gravitational field, which, in turn, is the vortex component of the total gravitational field

$$S^i(g,x^i) = \frac{^{\upsilon}\omega^i}{\ae}; \quad \ae = \frac{8\pi G}{c^4}. \quad (7)$$

Here the indices *(a), (b), (c), …* are local Lorentz indices, *i, k, l, m, ...* are world indices, and the tetrad coefficients $e^k_{(a)}$ satisfy the known orthonormality conditions:

$$e^k_{(a)}e^{(b)}_k = \delta^b_a; \quad e^k_{(a)}e^{(a)}_i = \delta^k_i; \quad e^{(a)}_i e_{k(a)} = g_{ik}; \quad e^k_{(a)}e_{k(b)} = \eta_{ab}. \quad (8)$$

Here $g_{ik}$ are the components of the metric space-time tensor, and $\eta_{ab}$ are the components of the metric tensor of the Minkowski tangent space.

Now, choosing the 4-velocity vector $\tau^i$ in formula (3) as the 4th timelike vector of the tetrad $e^i_{(4)}$, $e^i_4 = \tau^i$, we find the angular velocity of rotation of the vortex gravitational field from (6)

$$^{\upsilon}\omega^i = \frac{\lambda}{2\sqrt{k+1}}\delta^i_3, \quad (9)$$

which coincides with the angular velocity of rotation of a homogeneous space with metric (1):

$$^{\upsilon}\omega^i = \omega^i, \quad (10)$$

and represents the angular velocity of rotation of the congruences of time lines with the choice of the time-like tetrad vector $e^i_4 = \tau^i$.

Thus, it can be argued that the presence of a homogeneous vortex gravitational field and the rotation of a homogeneous space-time determine each other.

As mentioned above, spinning particles moving in the external gravitational field of a homogeneous space with rotation (1) will be described by the Dirac equation, which in the general covariant form has the form:

$$\gamma^k \nabla_k \psi + \mu\psi = 0; \quad \nabla_k \overline{\psi}\gamma^k - \mu\overline{\psi} = 0. \quad (11)$$

Here, $\psi(x^i)$ is the Dirac spinor function (bispinor), and $\overline{\psi}$ is the Dirac conjugate spinor function, and $\gamma_i$ are the Dirac matrices of the Riemannian space satisfying the condition of the fundamental connection between space and spin

$$\gamma_i\gamma_k + \gamma_k\gamma_i = 2g_{ik}I. \quad (12)$$

Here, $g_{ik}$ are the components of the metric tensor, and *I* is the identity 4-row matrix.

Using the tetrad formalism, the Dirac matrix $\gamma_i$ of a Riemannian space can be defined by the formulas:

$$\gamma_i = e^{(a)}_i \overset{0}{\gamma}_a; \quad \gamma^k = e^k_{(a)} \overset{0}{\gamma}{}^a, \quad (13)$$

where $e^k_{(a)}$ are the tetrad coefficients and $\overset{0}{\gamma}_a$ are the Dirac matrices of the Minkowski space satisfying the condition:

$$\overset{0}{\gamma}_i \overset{0}{\gamma}_k + \overset{0}{\gamma}_k \overset{0}{\gamma}_i = 2\eta_{ik}I, \quad (14)$$

where $\eta_{ik}$ are the components of the metric tensor of the Minkowski space.

It is easy to show that, taking into account relations (8) for tetrad coefficients $e_i^{(a)}$, condition (12) for Dirac matrices $\gamma_i$ is satisfied identically.

Knowing the metric coefficients in the metric (1), using formulas (13) we find the Dirac matrices in space-time (1):

$$\gamma_1 = \overset{0}{\gamma_1}; \quad \gamma_2 = \sqrt{k+1}\,e^{\lambda x}\overset{0}{\gamma_2} - e^{\lambda x}\overset{0}{\gamma_4}; \quad \gamma_3 = \overset{0}{\gamma_3};$$

$$\gamma^1 = \overset{0}{\gamma_1}; \quad \gamma^2 = \frac{e^{-\lambda x}}{\sqrt{k+1}}\overset{0}{\gamma_2}; \quad \gamma^3 = \overset{0}{\gamma_3}; \quad \gamma^4 = \frac{1}{\sqrt{k+1}}\overset{0}{\gamma_2} - \overset{0}{\gamma_4}. \qquad (15)$$

In the Dirac equation (11) $\nabla_k \psi$ and $\nabla_k \overline{\psi}$ are the covariant derivatives of the spinor functions $\psi$ and $\overline{\psi}$, which, using the usual postulate conditions for covariant differentiation (such as linearity, reality, covariance, the fulfillment of the Newton-Leibniz rule, etc.), can be written as

$$\nabla_k \psi = \partial_k \psi - \Gamma_k \psi; \quad \nabla_k \overline{\psi} = \partial_k \overline{\psi} + \overline{\psi}\Gamma_k. \qquad (16)$$

Here, $\Gamma_k$ are the coefficients of the spinor connection of the affine-metric space, which, using the tetrad formalism, are calculated by the formula [7, 8]:

$$\Gamma_k = -\frac{1}{4}\gamma^m\left(\partial_k \gamma_m - \Gamma_{km}^i \gamma_i\right), \qquad (17)$$

where $\Gamma_{km}^i$ are the connection coefficients of the affine-metric space. In the special case of a Riemannian space, the connection coefficients $\Gamma_{km}^i$ turn into Christophel symbols $\left\{\begin{array}{c}i\\km\end{array}\right\}$.

Now, using expressions (15) for matrices $\gamma_i$, by formula (17) we calculate the spinor connection coefficients in space-time with metric (1):

$$\Gamma_1 = \frac{\lambda}{4\sqrt{k+1}}\overset{0}{\gamma_1}\overset{0}{\gamma_2}; \quad \Gamma_2 = \frac{\lambda e^{\lambda x}(2k+1)}{4\sqrt{k+1}}\overset{0}{\gamma_2}\overset{0}{\gamma_4} - \frac{\lambda e^{\lambda x}}{4}\overset{0}{\gamma_1}\overset{0}{\gamma_2}; \quad \Gamma_3 = 0; \quad \Gamma_4 = \frac{\lambda}{4\sqrt{k+1}}\overset{0}{\gamma_1}\overset{0}{\gamma_2}. \qquad (18)$$

We consider that in our case the spinor function $\psi$ for a spinning particle depends only on time $t$ and does not depend on the spatial coordinates $(x, y, z)$, since the particle under consideration is in a homogeneous space-time, that is, $\psi = \psi(t)$.

Taking this into account and using formulas (15) and (18) for the matrices $\gamma_k$ and $\Gamma_k$, we write the Dirac equation (11) for the spinor functions $\psi(t)$ and $\overline{\psi}(t)$ in the form:

$$\left(\frac{1}{\sqrt{k+1}}\overset{0}{\gamma_2} - \overset{0}{\gamma_4}\right)\frac{d\psi}{dt} - \frac{\omega}{2}\left(\overset{0}{\gamma_3}\gamma_5\psi\right) + \frac{\lambda}{2}\overset{0}{\gamma_1}\psi + \mu\psi = 0;$$

$$\frac{d\overline{\psi}}{dt}\left(\frac{1}{\sqrt{k+1}}\overset{0}{\gamma_2} - \overset{0}{\gamma_4}\right) + \frac{\omega}{2}\left(\overline{\psi}\overset{0}{\gamma_3}\gamma_5\right) + \frac{\lambda}{2}\overline{\psi}\overset{0}{\gamma_1} - \mu\overline{\psi} = 0. \qquad (19)$$

Here $\gamma_5$ is the Dirac matrix $\gamma_5 = \overset{0}{\gamma_1}\overset{0}{\gamma_2}\overset{0}{\gamma_3}\overset{0}{\gamma_4}; \quad \gamma_5^2 = -1$.

Next, we will use equations (19) to determine the "observed" quantities of the spinor field. We will use composed quadratic combinations of spinor functions $\psi$, $\overline{\psi}$ and matrices $\gamma_i$ as "oserved" quantities, so that we will not need to find wave functions from (19). Such observable quantities are, for example, a scalar $\overline{\psi}\psi$ and a pseudoscalar $\overline{\psi}\gamma_5\psi$. This choice is due to the main components of the energy-momentum tensor of the spinor field (timelike vector of the current density $J_i$ of the spinor field $J_i = \overline{\psi}\overset{0}{\gamma_i}\psi$, and the axial spacelike vector of the flux density of the intrinsic angular momentum (spin) $S_i = \dfrac{\hbar c}{2}\left(\overline{\psi}\overset{0}{\gamma_i}\gamma_5\psi\right)$) are expressed through them.

As mentioned above, we use the comoving frame of reference for the spinor particles and the observer (3), and now, taking into account the spinor identity - the orthogonality of the current vector $\overline{\psi}\overset{0}{\gamma_i}\psi$ and the spin vector $\dfrac{\hbar c}{2}\left(\overline{\psi}\overset{0}{\gamma_i}\gamma_5\psi\right)$ [9, 10]

$$\left(\overline{\psi}\overset{0}{\gamma^i}\gamma_5\psi\right)\left(\overline{\psi}\overset{0}{\gamma_i}\psi\right) \equiv 0, \quad (20)$$

we obtain that in the comoving reference frame the time component of the spin vector $S_4 = \dfrac{\hbar c}{2}\left(\overline{\psi}\overset{0}{\gamma_4}\gamma_5\psi\right) = 0$, that is, the vector $S_i$ has only three space-like components, and it is located completely in a 3-dimensional spatial section and is a space vector

$$S_i = (S_x, S_y, S_z, 0) = \vec{S}. \quad (21)$$

Taking into account formula (21) for $S_i$ and forming various linear combinations of equations (19) for the spinor $\psi$ and the conjugate spinor $\overline{\psi}$ with each other, we obtain a system of differential equations for the axial spin vector $S_i$ and pseudoscalar $\overline{\psi}\gamma_5\psi$:

1) $\dfrac{d}{dt}(S_z) = 0$, or $S_z = \text{const}$;

2) $k\dfrac{d}{dt}(S_x) + 3(k+1)\omega(S_y) = 0$;

3) $\dfrac{d}{dt}(S_y) - \omega(S_x) = 0$; $\omega = \dfrac{\lambda}{2\sqrt{k+1}}$;

4) $\dfrac{1}{\sqrt{k+1}}\dfrac{d}{dt}(S_y) + \lambda(S_x) - 2\mu\left(\overline{\psi}\gamma_5\psi\right) = 0$;

5) $k\dfrac{d}{dt}\left(\overline{\psi}\gamma_5\psi\right) + 2\mu\sqrt{k+1}(S_y) = 0$.

$$(22)$$

It immediately follows from system (22) that the causality parameter $k$ cannot be equal to zero, since otherwise for $k=0$ all components of the vector $\overline{\psi}\overset{0}{\gamma_i}\gamma_5\psi = 0$, and pseudo-scalar $\overline{\psi}\gamma_5\psi = 0$, that is, all physical characteristics of the spinor particle vanish. Therefore, for the parameter k, two ranges of change are possible: $-1 < k < 0$, $k > 0$.

Further, from the first equation of system (22) it immediately follows that the projection of the spin vector $S_i = \dfrac{\hbar c}{2}\left(\overline{\psi}\overset{0}{\gamma_i}\gamma_5\psi\right)$ onto the rotation axis OZ is constant

$$S_z = \dfrac{\hbar c}{2}\left(\overline{\psi}\overset{0}{\gamma_3}\gamma_5\psi\right) = \text{const}, \quad (23)$$

and the 2nd and 3rd equations of this system at $k > 0$ taking into account relation (23) describe the process of precession of the spin of a Dirac spinor particle around the OZ axis with an angular velocity

$$\Omega = \omega\sqrt{\dfrac{3(k+1)}{k}}, \quad (24)$$

where $\omega = \dfrac{\lambda}{2\sqrt{k+1}}$ is the angular velocity of space rotation with metric (1) or, as shown above, the angular velocity of rotation of the vortex gravitational field. In this case, the angular velocity of precession for all admissible values of the parameter $k$ is greater than the angular velocity of space rotation $\omega$, and each of the components of the spin vector $S_x$ and $S_y$ satisfies the same equation of free harmonic oscillations with a frequency $\Omega$

$$\dfrac{d^2 S_x}{dt^2} + \Omega^2 S_x = 0 \;;\; \dfrac{d^2 S_y}{dt^2} + \Omega^2 S_y = 0 \quad (25)$$

Now, since $S_z = \text{const}$ if the origin of the vector $\vec{S}$ is located at the origin of coordinates $(0, 0, 0)$, then the end of this vector will always move in the same plane $(x, y)$, describing in it the corresponding curve $x = x(t)$, $y = y(t)$, where the functions $x(t)$ and $y(t)$ satisfy the same equation (25): $\dfrac{d^2 x}{dt^2} + \Omega^2 x = 0 \;;\; \dfrac{d^2 y}{dt^2} + \Omega^2 y = 0$. The solution of this system is :

$$x(t) = x_0 \cos(\Omega t) \;;\; y(t) = \dfrac{V_0}{\Omega} \sin(\Omega t) \quad (26)$$

Here, the initial conditions are accepted: $x(0) = x_0$, $y(0) = 0$, $\dfrac{dx}{dt}(0) = 0$, $\dfrac{dy}{dt}(0) = V_0$, where $x_0 = \sqrt{|\vec{S}|^2 - S_z^2}$.

Eliminating time from equations (26), we obtain the equation of the curve that describes the end of the vector $\vec{S}$ :

$$\dfrac{x^2}{x_0^2} + \dfrac{y^2}{(V_0/\Omega)^2} = 1 \quad (27)$$

In the general case, when $x_0 \neq V_0/\Omega$, equation (27) is the equation of an ellipse, that is, there will be an elliptical spin $\vec{S}$ precession, but in a particular case at $x_0 = V_0/\Omega$ we will have a circular precession.

For the opposite case ($k < 0$) with the same approach to describing the process of motion of the vector $\vec{S}$ as done above, when $k > 0$, and under the same initial conditions for the coordinates of the end of the vector $\vec{S}$ $x(t)$ and $y(t)$ we obtain solutions:

$$x(t) = x_0 \text{ch}(\Omega t) \;;\; y(t) = \dfrac{V_0}{\Omega} \text{sh}(\Omega t) \;;\; \Omega = \sqrt{\dfrac{3(k+1)}{|k|}} \omega \quad (28)$$

and, as result :

$$\dfrac{x^2}{x_0^2} - \dfrac{y^2}{(V_0/\Omega)^2} = 1 \quad (29)$$

That is the equation of a hyperbola. This kind of vector $\vec{S}$ motion can be called "hyperbolic precession".

Considering the system of equations (22), it is clear that we have 5 equations for four unknowns: $S_x$, $S_y$, $S_z$ and $\overline{\psi}\gamma_5\psi$, that is, we have an overdetermined system of equations.

From the compatibility conditions for the system of equations (22) for the axial spin vector $S_i$ and pseudoscalar $\overline{\psi}\gamma_5\psi$ of the Dirac spinning particle, the relation between the parameters characterizing the geometry of the considered space-time λ, k, ω, and the parameters of the Dirac particles *μ, m,* where *m* is their mass, follows:

$$\sqrt{\frac{3}{4}(2k+3)}\,\omega = \mu c \; ; \quad \mu = mc/\hbar . \quad (30)$$

It should be mentioned that in this formula the parameter ω is the angular velocity of spatial rotation and at the same time the angular velocity of rotation of congruences of time-like curves (for example, time-like geodesics), that is, the kinematic characteristic of the vortex gravitational field.

Further, it will be interesting for us to consider the option when the causality parameter $k<0$, moreover, when $-1 < k \leq -1/2$. In this case, the factor $\sqrt{\frac{3}{4}(2k+3)}$ differs very little from unity, and for $k = -5/6$ it is exactly equal to unity. Therefore, taking into account what has been said, the compatibility condition (30) can be rewritten in the form

$$\omega = \frac{mc^2}{\hbar} . \quad (31)$$

On the other hand, if we consider the proper angular momentum (spin) of a Dirac particle, such as an electron, at the classical level, assuming that its spin $S = \hbar/2$ due to the rotation of the particle around the axis, and its dimensions are equal to the reduced Compton wavelength $\lambda_C = \frac{\hbar}{mc}$, then we will have:

$$I\omega = \frac{\hbar}{2} ; \quad \omega = \frac{\hbar}{2I} . \quad (32)$$

Here *I* is the moment of inertia of the particle about the axis of rotation. Therefore, assuming that the electron has the shape of a cylinder with radius $\lambda_C = \frac{\hbar}{mc}$, and $I = \frac{1}{2}m\left(\frac{\hbar}{mc}\right)^2$, from formula (32) we will have:

$$\omega = \frac{\hbar}{m}\frac{m^2 c^2}{\hbar^2} = \frac{mc^2}{\hbar} . \quad (33)$$

We have obtained complete agreement with formula (31) for the angular velocity of rotation of congruences of timelike geodesics.

This means that the electric substance of the electron, through the points of which the congruence of timelike geodesics also passes, is also a part of the gravitational vortex field, or more precisely, that the spin of the electron induces its own gravitational vortex.

This is also indicated by the general theory of the gravitational interaction of the spinor field. It is known that the interaction of the spinor and gravitational fields is described by the Lagrangian [6]

$$L_{\text{int}}(\psi, g) = \frac{\hbar c}{2}\omega^i \left(\overline{\psi}\gamma_i \gamma_5 \psi\right), \quad (34)$$

where $\omega^i = \frac{1}{2}\varepsilon^{iklm} e_{k(a)} e^{(a)}_{l,m}$ is the angular velocity of rotation of the tangent tetrad, as mentioned above, is the kinematic characteristic of the vortex gravitational field, and $\frac{\hbar c}{2}(\bar{\psi}\gamma_i\gamma_5\psi)$ is the spin flux density vector of the spinor field.

In addition, formulas (31) and (33) can also be written in the following form:

$$\hbar\omega = mc^2. \quad (35)$$

Here, the left side of the formula represents the energy of the rotational motion of the electron due to its spin, and the right side ($mc^2$) is the rest energy of the electron, and $m$ is its rest mass. Therefore, formula (35) can be interpreted in such a way that the rotation of the electron, that is, its spin, induces the appearance of the electron rest mass, or, more simply, that the electron spin determines the existence of the electron rest mass.

There is another important aspect in solving the problem of the motion of a Dirac particle in space with rotation for the case of negative values of the causality parameter $k < 0$. As it has been pointed out more than once for $k < 0$ at least one time-like geodesic passes through each point of space-time, describing the motion of a particle, that is, such a curve, leaving the point $y = 0$, $t = 0$ according to the observer's clock, will return back at time $t = 0$, but, as numerical studies show, presented graphically in fig. 1, at another point in the XOY plane at a distance $\Delta y$ from the origin, that is, the observer will see the same particle at the same time, but with uncertainty in its location $\Delta y$. Since the momentum of the particle $p_y = mV$ is not precisely defined, that is, its uncertainty $\Delta p_y = mV$, and, as numerical calculations show, the value $\Delta y \sim \frac{\hbar}{mV} \sim \frac{\hbar}{\Delta p_y}$, that is, of the order of the de Broglie wavelength of the spinor particle, from here we obtain $\Delta y \Delta p_y \sim \hbar$. Here we get in a geometric interpretation the uncertainty principle of quantum mechanics. In addition, as shown in Fig. 1, the same geodesic at the moment of time $t = 0$ will once again intersect the XOY plane, but at a more distant point $\Delta y_1$ ($\Delta y_1 > \Delta y$), it turns out that the particle is, as it were, spread out in the interval ($\Delta y$, $\Delta y_1$).

However, the observer himself moves in time along the *Ot* axis, remaining at the point $y = 0$, and signals that the particle is at the same time $t = 0$ at a distance $\Delta y$ and then (according to the observer's clock) at a distance $\Delta y_1$ will reach him when he moves along the axis *Ot* to time intervals $\Delta t$ and $\Delta t_1$ in accordance with the transit time of these signals ($\Delta t < \Delta t_1$).

It turns out that the "smearing" of the spinning particle occurs as the real time grows, the time of the observer, and the "smearing" of this particle as the real time grows can be interpreted as the spreading of the wave packet of this particle.

As a result, we have obtained a geometric interpretation of the uncertainty principle in quantum mechanics and the spreading of the wave packet of microparticles in the framework of general relativity.

Summarizing the results of this work, the following conclusions can be drawn:
1) 1) When spinning particles move, obeying the Dirac equation, in a homogeneous space with rotation, that is, in a uniform vortex gravitational field, the angular momentum (spin), equal to $\hbar/2$, will precess around the axis of rotation of the gravitational vortex.
2) If a Dirac particle is considered as a rotating cylinder with a radius equal to its Compton wavelength $\frac{\hbar}{mc}$, around its axis parallel to the axis of rotation of the vortex gravitational field, and the angular momentum of this particle (spin), equal to $\hbar/2$, is considered as a consequence of such rotation, then we obtain that the frequency of its rotation is exactly equal to the rotation frequency ω of the vortex gravitational field, in the space of which the particle moves.

3) It is shown that the inertial mass *m* of a Dirac particle, such as an electron, is induced by its rotation, that is, by spin, in accordance with the formula $m = \hbar\omega/c^2$, that is, the inertial mass and spin are interconnected and inseparable.

4) The space-time in the nearest vicinity of fermions (of the order of several wavelengths $\frac{h}{mV}$ of de Broglie particles ) is the space-time described by the metric (1) with a uniform vortex gravitational field induced by the fermion spin.

5) A geometric interpretation of the uncertainty principle of quantum mechanics and the wave packet spreading effect was obtained at the classical level within the framework of general relativity.